\documentclass[final,3p,times,twocolumn]{elsarticle}

\usepackage{graphicx}

\usepackage{amssymb}

\usepackage{lineno}

\newcommand{\rmi}{\mathrm{i}}

\journal{Physica C}

\begin{document}

\begin{frontmatter}



\title{Angular dependence of the high-frequency vortex response in YBa$_2$Cu$_3$O$_{7-x}$ thin film with self-assembled BaZrO$_3$ nanorods}


\author[RomaTre]{N. Pompeo\corref{cor1}}
\cortext[cor1]{Corresponding author.}
\ead{pompeo@fis.uniroma3.it}
\author[RomaTre]{R. Rogai}
\author[RomaTre]{K. Torokhtii}
\author[ENEA]{A. Augieri}
\author[ENEA]{G. Celentano}
\author[ENEA]{V. Galluzzi}
\author[RomaTre]{E. Silva}
\address[RomaTre]{Dipartimento di Fisica ``E. Amaldi'' and Unit\`a CNISM,
Universit\`a Roma Tre, Via della Vasca Navale 84, 00146 Roma,
Italy}

\address[ENEA]{ENEA-Frascati, Via Enrico Fermi 45, 00044 Frascati, Roma, Italy}

\begin{abstract}
We present a microwave study of the angular dependence of the flux-flow resistivity $\rho_{ff}$ and of the pinning constant $k_p$ in YBCO thin films containing BZO nanorods. We find that BZO nanorods are very efficient pinning centers, even in tilted fields. We find that $\rho_{ff}$ is a scaling function of a reduced field $H/f(\theta)$. We extend a model for the anisotropic motion of vortices in uniaxially anisotropic superconductor, able to describe the experimental $f(\theta)$ on the basis of only the intrinsic anisotropy of YBCO. The pinning constant $k_p$, by contrast, exhibits different field dependences in different angular ranges, consistent with pinning by BZO at angles as large as 60$^{\circ}$, and with pinning along the $a,b$ planes as originating from the same mechanism as in pure YBCO with the field along the $c$ axis.
\end{abstract}


\end{frontmatter}


\section{Introduction and model}
\label{intro}
Enhanced pinning by artificial defects, such as those originated by BaZrO$_3$ (BZO) inclusions within YBa$_2$Cu$_3$O$_{7-x}$ (YBCO) films, has been extensively studied in the last years due to the potential impact on applications, with substantially increased critical currents and irreversibility fields \cite{macmanusNATMAT04,kangSCI06,gutierrezNATMAT07}. Issues connected to the directional nature of the pinning efficiency have been under careful scrutiny, since BZO defects self-assemble typically in linear structures roughly parallel to the $c$ axis, thus competing with intrinsic pinning due to the layered structure itself. As a consequence, the angular dependence of critical current density $j_c$ is heavily affected by BZO defects \cite{macmanusNATMAT04,augieriIEEE09}, with a maximum of $j_c$ along the $c$ axis. Whether this behaviour competes or coexists with the intrinsic anisotropy of YBCO is difficult to ascertain with a dc technique, where the pinning potential is probed at high current densities, that is when the flux lines can be extracted from the pinning potential wells and moved apart.

We present in this paper a microwave study of the angular dependence of the vortex motion resistivity in a YBZO/BZO film, where we disentangle the effects of directional pinning from the intrinsic anisotropic resistivity, exploiting the peculiarities of the microwave complex response. At microwave frequencies the flux lines oscillate over distances of less than $\sim 1$ nm \cite{tomaschPRB88}, thus probing only the bottom (i.e., the steepness) of the pinning potential. Interestingly, we observed previously a strong increase of pinning as measured at $\sim$50 GHz in YBCO/BZO films with different BZO content \cite{pompeoAPL07}, thus showing that BZO induces very steep pinning wells. The short-distance oscillations allow in most cases the application of mean-field theories developed for the vortex motion resistivity $\rho_{vm}$ \cite{GR,CC,brandtPRL91}. Comparison with the data is made easier by the fact that those theories can be described by a unified formulation \cite{pompeoPRB08}:
\begin{equation}
\label{eq:rhovm}
    \rho_{vm}=\rho_{vm,1}+\rmi\rho_{vm,2}=\rho_{ff}\frac{\varepsilon+\rmi\left(\nu/\bar{\nu}\right)}{1+\rmi\left(\nu/\bar{\nu}\right)}
\end{equation}
\noindent where the resistivity $\rho_{ff}$ represents the genuine flux flow resistivity (only viscous drag), reached at high frequency instead of at high currents, $\bar{\nu}$ is a characteristic frequency and the dimensionless parameter $0\leq\varepsilon\leq 1$ is a measure of thermal activation phenomena. When $\varepsilon$ is small, the model reduces to the well-known Gittleman-Rosenblum model \cite{GR} and $\bar{\nu}\rightarrow k_p \rho_{ff}/\Phi_0 B$. A complete discussion on the extraction of the vortex parameters $k_p$ (pinning constant) and $\rho_{ff}$, together with the assessment of the validity limits, has been given in \cite{pompeoPRB08}. In the measurements of the angular and field dependence of $\rho_{ff}$ and $k_p$  here reported, the systematic (not scattering) error is less than 20\%.

The flux-flow resistivity, as an intrinsic property of superconductors, is a model-dependent function \cite{LO,troyPRB93,BS} of the reduced field $h=H/H_{c2}$ (in the London approximation, where $B\simeq\mu_0 H$), $\rho_{ff}=\rho_n F(h)$, where $\rho_n$ is the normal state resistivity. The Bardeen-Stephen (BS) model \cite{BS} predicts linearity with the field, $\rho_{ff}=\rho_n h$. When $H$ is tilted by an angle $\theta$ from the $c$ axis, $H_{c2}$ acquires an angular dependence: $H_{c2}(\theta)=H_{c2}(0) \varepsilon(\theta)$, with $\epsilon(\theta)=\left[\cos^2\theta+\gamma^{-2}\sin^2\theta\right]^{1/2}$ in the anisotropic 3D Ginzburg-Landau model, with $\gamma=H_{c2}(0^{\circ})/H_{c2}(90^{\circ})\simeq 5\div 8$ in YBCO. The resulting flux flow resistivity becomes
\begin{equation}
\label{eq:rhoff}
    \rho_{ff}(H,\theta)=\rho_n {H}/{H_{c2}(0^{\circ})}\varepsilon(\theta)
\end{equation}
In dc, Eq. \ref{eq:rhoff} holds for the genuine flux flow resistivity, and when the force acting on flux lines does not change with the angle: this is typically the case when pinning is irrelevant or isotropic (e.g. in the crossover region between the flux-flow and fluctuation regimes \cite{sartiPRB97}) and the Lorentz force on flux lines is constant. When directional pinning plays a role, the scaling hypothesis for the measured resistivity breaks down: a scaling ($\rho_{ff}$) and a non-scaling (pinning force) quantities mix to give the observed total vortex motion resistivity $\rho_{vm} \neq \rho_{ff}$, so that a comprehensive description of the experiments becomes difficult.

A second issue, relevant to this paper, comes when the angle between $H$ and the current is varied. In this case the driving force changes with the angle.  On a theoretical ground, this is far from being a trivial problem, and a general expression for $\rho_{ff}$ can be written only under restrictive assumptions. Using a time-dependent Ginzburg Landau approach, Hao, Hu and Ting (HHT) \cite{HHTPRB52} have obtained an expression for the flux flow resistivity in anisotropic superconductors. In spherical coordinates, where $c \parallel z$, $\theta$ is the angle of $H$ with $z$ and $\phi$ is the angle between $x$ and the in-plane projection of $H$, in a uniaxial superconductor with $J \parallel x$ we reduce the HHT expression to:
\begin{equation}
\label{eq:rhox}
    \rho_{ff}(\theta,\phi)=\rho_{ab}(\theta)\frac{\rho_{ab}(\theta) \sin^2\theta\sin^2\phi+\rho_{c}(\theta)\cos^2\theta}{\rho_{ab}(\theta) \sin^2\theta+\rho_{c}(\theta)\cos^2\theta}
\end{equation}
where $\rho_{ab}, \rho_{c}$ are the intrinsic $a,b$ plane and $c$ axis flux flow resistivity,\footnote{While $\rho_{ab}(\theta)$ can be directly measured when $\phi=0$, see Eq.(\ref{eq:rhox}), $\rho_{c}(\theta)$ can be derived (see  \cite{HHTPRB52} for extensive discussion) only with a multiterminal technique (such as, e.g., in \cite{espositoJAP00}).}
and $\rho_{ff}$ stands for the experimentally measured, in-plane flux flow resistivity. When (i) the anisotropic scaling applies, (ii) the flux flow resistivities are linear with the field (BS model), and (iii) $\rho_{c,n}/\rho_{ab,n}\simeq \gamma^2$, we obtain
\begin{equation}
\label{eq:localrhop}
    \rho_{ff}(\theta,\phi)=\rho_{ab}(\theta=0)\frac{\gamma^{-2}\sin^2\theta\sin^2\phi+\cos^2\theta}{(\gamma^{-2}\sin^2\theta+\cos^2\theta)^{1/2}}
\end{equation}
Eq. (\ref{eq:localrhop}) is obtained under strong restrictions (in particular, (iii) is only approximate, even if the order of magnitude is correct \cite{okuyaJS94,odaPRB88}). Nonetheless, in nearly optimally doped or overdoped YBCO we do not expect significant deviations from the assumptions (i)-(iii). In Sec. \ref{results} we will compare an extension of Eq. (\ref{eq:localrhop}) to our anisotropic data for $\rho_{ff}$.
\section{Experimental section}
\label{Exp}
\noindent
We hereby analyze a YBCO/BZO sample grown by PLD from targets containing BaZrO$_3$ (BZO) powders at 5\% mol. (details are extensively given in \cite{galluzziIEEE07}). The BZO inclusions generated columnar-like defects, approximately perpendicular to the film plane, as observed with transverse TEM images \cite{augieriJAP10}. The directionality of the defects is consistent with the angular dependence of the critical current density \cite{augieriIEEE09}.
\begin{figure}[h]
\centerline{\includegraphics[width=8.5cm]{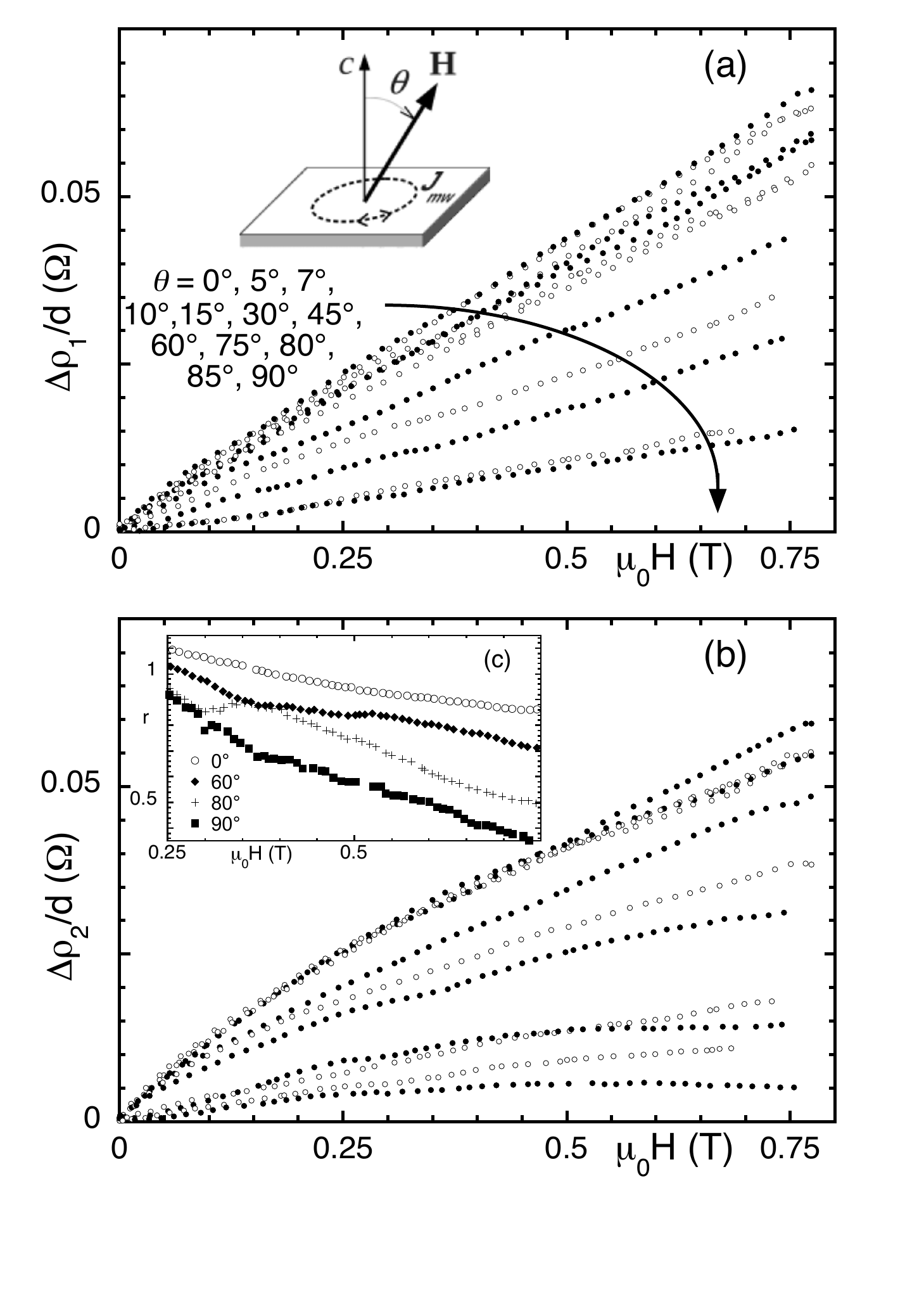}}
\vspace{-10mm}
  \caption{Field induced variation of the real (a) and imaginary (b) microwave resistivity at $T=81$ K and different angles $\theta$ between the dc field and the $c$ axis. For the sake of readability, only a subset of the data are reported. The inset of panel (a) sketches the microwave current pattern and the definition of the angle $\theta$. (c) Field dependence of $r=\Delta\rho_2/\Delta\rho_1$ at selected angles and expanded field scale.}
\label{figraw}
\end{figure}

The complex microwave effective surface impedance has been measured by a sapphire cylindrical dielectric resonator operating at $\sim$48 GHz, extensively described elsewhere \cite{pompeoJS07}, so that we report here only the results. For the discussion of the data, it is important to stress that the microwave currents $J_{mw}$ flow parallel to the film plane (i.e. parallel to the $a,b$ planes) on a circular fashion (see inset of Fig. \ref{figraw}). A dc magnetic field $\mu_0 H\leq$0.8 T was applied at the angle $\theta$ with the $c$ axis. Field sweeps were performed at different angles. We verified that the angular response was symmetric with respect to the $a,b$ planes. The raw data yielded $\Delta\rho(H)/d=\left[ \rho (H)-\rho (0) \right]/d=\left[ \Delta\rho_1 + \mathrm{i}\Delta\rho_2 \right]/d$, where $\rho=\rho_1+\mathrm{i}\rho_2$ is the complex resistivity and $d$ is the film thickness. When $T$ is sufficiently below $T_c$ one can neglect the field-induced pair-breaking. Thus, the subtraction of the zero-field value allows one to identify $\Delta\rho$ with $\rho_{vm}$. Taking into account this requirement and the disappearance of the signal in the noise as $\theta \rightarrow 90^{\circ}$ at low $T$, measurements were performed at $T=81$ K as a compromise to obtain a reasonable signal for all orientations. 

Fig. \ref{figraw} reports the field dependence of $\Delta\rho_1/d$ and $\Delta\rho_2/d$ for several angles $\theta$. As expected, $\Delta\rho(H)$ decreases with increasing angle, due to the intrinsic anisotropy of YBCO and to the reduced effect of the Lorentz force on flux lines (when $\theta=0$, $H \perp J_{mw}$, while when $\theta=90^{\circ}$ only part of the current lines are perpendicular to the field, thus reducing the net effective Lorentz force). The ratio $r=\Delta\rho_2 / \Delta\rho_1 $ expresses an experimental measure of the balance between reactive and dissipative contributions to the response. Roughly speaking,  $r>1$ indicates strong pinning. Fig. \ref{figraw}c shows that the overall response becomes progressively dominated by the dissipative contribution with $\theta\rightarrow 90^{\circ}$. This interesting feature, that might suggest peculiar angular dependence of pinning and of $\rho_{ff}$, requires the explicit dependence of the vortex parameters for a quantitative discussion. 
\section{Discussion}
\label{results}
\begin{figure}[h]
\centerline{\includegraphics[width=8cm]{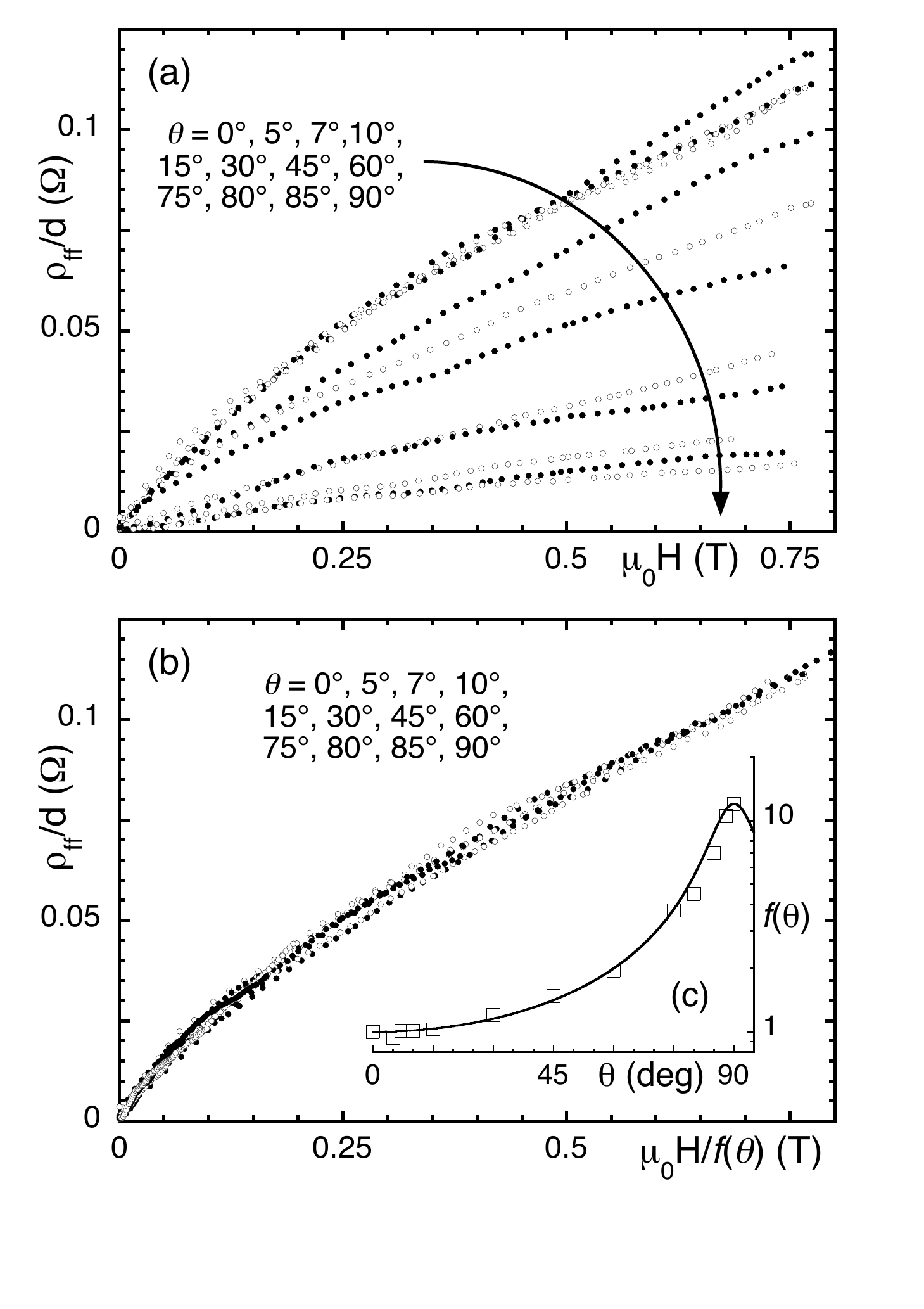}}
\vspace{-10mm}
  \caption{(a) $\rho_{ff}(H)$ at various $\theta$, derived from the measurements of $\Delta\rho$; only 10\% of the data have been plotted to avoid crowding. (b) Angular scaling $\rho_{ff}(H/f(\theta))$ for the data in (a). (c) Experimental scaling function $f(\theta)$ (squares) and the anisotropic scaling function corrected with the reduction of the Lorentz force (continuous line) with $\gamma=6$ (see text); the size of the squares is a measure of the error bars.}
\label{figrhoff}
\end{figure}

We first discuss $\rho_{ff}$. In Fig. \ref{figrhoff}a we report $\rho_{ff}$ at selected angles as derived from the data \cite{pompeoPRB08}. As a matter of fact, we are able to collapse all the curves (angular scaling) with an empirical angular scaling function $f(\theta)$. We report in Fig. \ref{figrhoff}b the collapsed curves of $\rho_{ff}$ vs $H/f(\theta)$ and the scaling function $f(\theta)$. As it can be seen, the anisotropy $f(90^{\circ})/f(0^{\circ})$=12 is significantly larger than the commonly accepted values for the intrinsic anisotropy $\gamma=5\div 8$. We ascribe this discrepancy to the varying Lorentz force with $\theta$. Since in our experiment the microwave currents have a circular pattern, we average Eq. (\ref{eq:localrhop}) over $\phi$ and we obtain
\begin{equation}
\label{eq:rhoeff}
    \langle\rho_{ff}(\theta,\phi)\rangle_{\phi}=\rho_{ab}(0^{\circ})\frac{\frac{1}{2}\gamma^{-2}\sin^2\theta+\cos^2\theta}{(\gamma^{-2}\sin^2\theta+\cos^2\theta)^{1/2}}
\end{equation}
Taking Eq. (\ref{eq:rhoff}) for $\rho_{ab}$, one finds again a scaling law, with $1/f(\theta)$ given by the fraction in Eq.(\ref{eq:rhoeff}). This angular function is plotted in Fig. \ref{figrhoff}c (solid line). The experimental datum point $f(90^{\circ})=12$ fixes $\gamma=6$, within the range of commonly accepted values. The zero-parameter curve describes the data well. In particular, it recovers the large effective anisotropy. We can thus state that the flux-flow resistivity, as expected, is an intrinsic property dictated only by the electronic anisotropy of YBCO.

\begin{figure}[h]
\centerline{\includegraphics[width=8.cm]{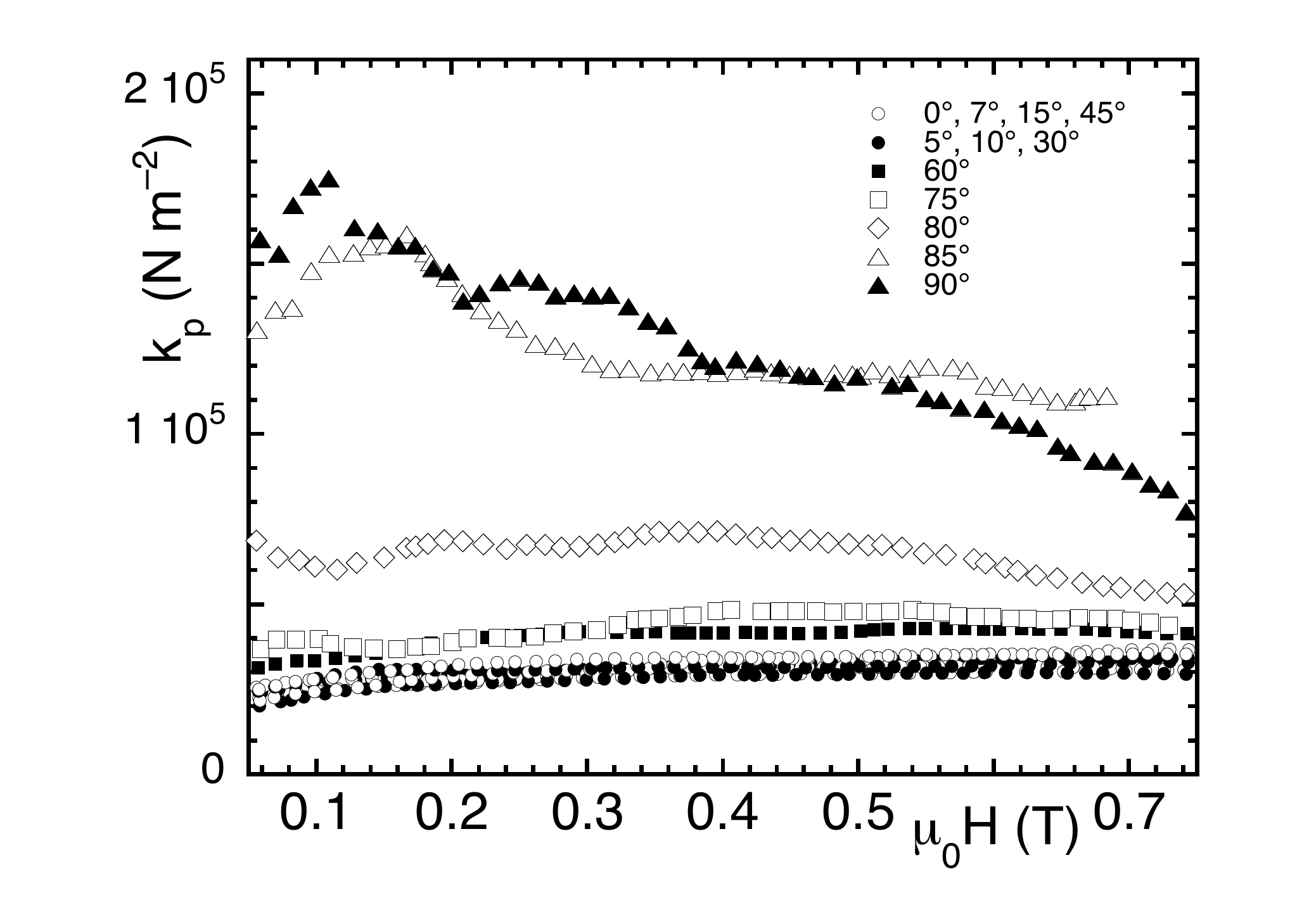}}
\vspace{-5mm}
  \caption{$k_p$ vs. $H$ for the same angles as in Fig. \ref{figrhoff}; data below $\mu_0H=$50 mT are unreliable due to the inversion procedure and are not reported. It is immediately seen that an angular scaling is impossible. Different field dependences, pointing to different pinning mechanisms, develop in different angular regions.}
\label{figkp}
\end{figure}

We finally comment on the field dependence of $k_p$. In Fig. \ref{figkp} we report $k_p$ vs. $H$ at various $\theta$. There is clearly no possibility to collapse the data onto a single curve by simply rescaling the field, as for $\rho_{ff}$. The field dependence of $k_p$ gives some insight. In particular, different pinning regimes set in, in different angular regions: in the wide region $0^{\circ}\leq\theta\leq45^{\circ}$, $k_p$ is basically constant (a close inspection reveals that $k_p$ increases slightly with the field) and, interestingly, it is almost angle-indipendent. With $60^{\circ}\leq\theta\leq80^{\circ}$ the field dependence remains the same, but the absolute values shift upward. With further increasing $\theta\geq85^{\circ}$, $k_p$ acquires a strong dependence with the field, decreasing quickly as $H$ increases.

Microwave measurements at $\theta=0^{\circ}$ showed \cite{pompeoAPL07} that a weakly field-increasing $k_p$ was a feature of YBCO/BZO films, as opposed to field-decreasing $k_p$, that was typical of pure YBCO. By assigning the constant $k_p$ to the effect of BZO, we get that in the present measurements the pinning mechanism is BZO dominated up to (at least) $45^{\circ}$. By contrast, since the field decrease of $k_p$ is typical of pure YBCO, on qualitative grounds the pinning mechanism in fields nearly parallel to the $a,b$ planes in YBCO/BZO looks the same as in pure YBCO, with the magnetic field along the $c$ axis. In this scenario, the highest absolute values of $k_p(90^{\circ})$ are presumably due mostly to the anisotropy, while for $60^{\circ}\leq\theta\leq85^{\circ}$ there is an interplay between pinning by BZO and by anisotropy.
\section{Summary}
\label{conc}
We have presented angular measurements of the genuine flux flow resistivity $\rho_{ff}$ and pinning constant $k_p$ in a YBCO film with BZO nanorods. By taking into account the variable Lorentz force in our experimental setup, we have shown that $\rho_{ff}(H,\theta)=\rho_{ff}(H/f(\theta))$, where only the intrinsic $H_{c2}$ anisotropy plays a role. By contrast, $k_p$ shows clear indications of two preferred directions for pinning, ascribed to BZO nanorods and to the $a,b$ planes. The information gained from high-frequency investigations are complementary to dc measurements, since the genuine $\rho_{ff}$ is accessible, and steepness of the pinning potential, rather than the depth, is probed.
\\
\\
We thank S. Schweizer for the help in taking data. This work has been partially supported by the FIRB project ``SURE:ARTYST" and by EURATOM. N.P. acknowledges support from Regione Lazio.






\end{document}